# Degradation mechanism of $CH_3NH_3PbI_3$ perovskite materials upon exposure to humid air

Masaki Shirayama,[1] Masato Kato,[1] Tetsuhiko Miyadera,[2] Takeshi Sugita,[2] Takemasa Fujiseki,[1] Shota Hara,[1] Hideyuki Kadowaki,[1] Daisuke Murata,[1] Masayuki Chikamatsu,[2] and Hiroyuki Fujiwara[1,a]

[1]Department of Electrical, Electronic and Computer Engineering, Gifu University, 1-1 Yanagido, Gifu 501-1193, Japan

[2]Research Center for Photovoltaics, National Institute of Advanced Industrial Science and Technology (AIST), Central 5, 1-1-1 Higashi, Tsukuba, Ibaraki 305-8568, Japan.

## Abstract

Low stability of organic-inorganic perovskite ($CH_3NH_3PbI_3$) solar cells in humid air environments is a serious drawback which could limit practical application of this material severely. In this study, from real-time spectroscopic ellipsometry characterization, the degradation mechanism of ultra-smooth $CH_3NH_3PbI_3$ layers prepared by a laser evaporation technique is studied. We present evidence that the $CH_3NH_3PbI_3$ degradation in humid air proceeds by two competing reactions of (i) the $PbI_2$ formation by the desorption of $CH_3NH_3I$ species and (ii) the generation of a $CH_3NH_3PbI_3$ hydrate phase by $H_2O$ incorporation. In particular, rapid phase change occurs in the near-surface region and the $CH_3NH_3PbI_3$ layer thickness reduces rapidly in the initial 1-h air exposure even at a low relative humidity of 40%. After the prolonged air exposure, the $CH_3NH_3PbI_3$ layer is converted completely to hexagonal platelet $PbI_2$/hydrate crystals that have a distinct atomic-scale multilayer structure with a period of 0.65 ± 0.05 nm. We find that conventional x-ray diffraction and optical characterization in the visible region, used commonly in earlier works, are quite insensitive to the surface phase change. Based on results obtained in this work, we discuss the degradation mechanism of $CH_3NH_3PbI_3$ in humid air.

[a]Author to whom correspondence should be addressed. Electronic mail: fujiwara@gifu-u.ac.jp.

## I. INTRODUCTION

In large-area photovoltaic devices, long-term stability is one of the critical factors as the cost of solar cell modules is essentially determined by the life cycle of the modules. Although quite high conversion efficiencies of ~19% have been reported for methylammonium lead iodide perovskite ($CH_3NH_3PbI_3$) solar cells,[1,2] this unique organic-inorganic hybrid material suffers from rather intense degradation in humid air.[2-23] In particular, the conversion efficiency of $CH_3NH_3PbI_3$ solar cells has been reported to reduce by more than 50% in periods of 1–2 days.[2,3] Many earlier studies confirmed or proposed that the degradation of the solar cells is triggered by $H_2O$ molecules in air.[2-23] Quite interestingly, a high efficiency of 19.3% can be obtained even when the perovskite device is made at a relative humidity (RH) of ~30%, although this device still shows rapid degradation behavior in humid air.[2] The characteristics of the hole transport layer also affect the degradation. Specifically, Li ions introduced as dopants in the hole transport layer deteriorate the $CH_3NH_3PbI_3$ stability significantly,[5,6,8] as the incorporation of Li dopants enhances the hygroscopic properties.

The long-term stability of $CH_3NH_3PbI_3$ solar cells can be achieved by incorporating the $H_2O$-barrier layers in the rear side of the solar cells.[4-7] For this purpose, C-based electrodes,[4,24] oligothiophene derivative,[5] poly(methylmethacrylate),[6] and atomic-layer-deposited $Al_2O_3$ (Ref. 7) have been applied successfully. Accordingly, it is essential to control the $H_2O$ permeation through a hole transport layer/rear electrode structure to prevent the intense $CH_3NH_3PbI_3$ degradation. In the solar cells with the $H_2O$-barrier layers, however, lower conversion efficiencies of 8.8–15.3% have been reported[4-7,24] and thus further improvement is necessary.

For the suppression of the $CH_3NH_3PbI_3$ degradation, the understanding of the degradation process is expected to be quite helpful. Numerous earlier studies conclude that $CH_3NH_3PbI_3$ degradation in humid air occurs mainly by the simple decomposition of $CH_3NH_3PbI_3$ into $PbI_2$ and $CH_3NH_3I$ (or $CH_3NH_2$ + HI), resulting in the $PbI_2$ phase precipitation by the desorption of volatile $CH_3NH_3I$-related species.[9-14] This decomposition reaction is considered to proceed by a strong interaction between $H_2O$ and a N−H bond in $CH_3NH_3PbI_3$, which leads to the decomposition of the $PbI_3^-$ cage with $CH_3NH_3^+$ inside.[6,7] In fact, after the $CH_3NH_3PbI_3$ degradation, the formation of the $PbI_2$ phase can be confirmed rather easily from x-ray diffraction (XRD).[10-12,14-16] The visible absorption spectroscopy also shows that the onset of the light absorption shifts up to 2.4 eV after the degradation,[8,10,11,16,17] which is consistent with the $PbI_2$ formation. Later, however, it was suggested that the reaction of $CH_3NH_3PbI_3$ with $H_2O$ leads to the formation of a hydrate phase, such as $(CH_3NH_3)_4PbI_6\cdot 2H_2O$, in addition to $PbI_2$.[18] More recently, Leguy *et al.* found that the crystal structure formed after the degradation is a different hydrate phase of $CH_3NH_3PbI_3\cdot H_2O$ and proposed that the intermediate $CH_3NH_3PbI_3\cdot H_2O$ further decomposes into $(CH_3NH_3)_4PbI_6\cdot 2H_2O$ and $PbI_2$ phases.[19] At this stage, however, reaction processes or mechanisms proposed for $CH_3NH_3PbI_3$ degradation in humid air are highly controversial[6-14,16,18,19] and further detail on the reaction pathway needs to be determined.

When the phase transformation of $CH_3NH_3PbI_3$ in humid air is characterized by optical spectroscopy, on the other hand, the strong loss of visible-light absorption occurs during the initial 1–5 days at high RH conditions (RH ≥ 80%),[8,18] whereas the degradation is quite slow at the moderate conditions (RH ≤ 60%) and, in this case, the structural change continues over a prolonged period (1 month).[12,15,18] In contrast, the degradation of $CH_3NH_3PbI_3$ solar cells at similar conditions (~50% RH) occurs in a much shorter time scale (1–2 days).[3] Accordingly, it is possible that the phase transformation is more significant in the near-surface (or interface) region, leading to the rapid reduction in the solar cell efficiencies. Nevertheless, solution-processed $CH_3NH_3PbI_3$ films have quite rough surface structures and these layers often exhibit highly non-uniform morphologies,[8,12,13,18] which make detailed near-surface characterization rather difficult.

In this study, we have characterized the degradation process of ultra-smooth $CH_3NH_3PbI_3$ layers in humid air using real-time spectroscopic ellipsometry (SE). To enhance the surface sensitivity in the measurements, quite thin $CH_3NH_3PbI_3$ layers (45 nm) are evaluated. We find that the phase change in the surface region occurs quite rapidly even at a low RH of 40% and the unique multilayer structure is formed after the complete degradation. Determination of the dielectric function ($\varepsilon = \varepsilon_1 - i\varepsilon_2$) in the UV region further allows us to identify the near-surface structure during the $CH_3NH_3PbI_3$ degradation in humid air. To understand the optical properties of the hydrate phases, density functional theory (DFT) calculations are further performed. Based on results obtained in this study, we propose that $CH_3NH_3PbI_3$ degradation in humid air proceeds by two competing formation reactions of $PbI_2$ and $CH_3NH_3PbI_3$ hydrate crystals.

## II. EXPERIMENT

For reliable SE analysis, preparation of samples with smooth surface is essential.[25] To prepare $CH_3NH_3PbI_3$ layers with ultra-smooth surfaces, we employed a laser evaporation technique.[26] In this process, $PbI_2$ and $CH_3NH_3I$ source materials in crucibles are heated by a near-infrared laser with a wavelength of $\lambda$ = 808 nm. The controllability of the $CH_3NH_3I$ evaporation rate is improved greatly in this technique, compared with conventional evaporation processes. In particular, the $CH_3NH_3I$ source has high vapor pressure and, to perform the precise control of the $CH_3NH_3I$ evaporation rate, the near-infrared laser is modulated by a 10 Hz square wave. Our evaporation process is conducted without substrate heating at a pressure of $5\times10^{-3}$ Pa. For the $CH_3NH_3PbI_3$ deposition, we used crystalline Si (*c*−Si) substrates coated with ZnO layers (50 nm) to improve film adhesion. To suppress surface roughening and structural non-uniformity, we characterized thin $CH_3NH_3PbI_3$ layers (45 nm). We emphasize that the structure and optical properties of the laser-evaporated $CH_3NH_3PbI_3$ layers are essentially the same with conventional evaporated and solution processed $CH_3NH_3PbI_3$ layers.[26]

To reveal the degradation mechanism of $CH_3NH_3PbI_3$, we exposed ultra-smooth $CH_3NH_3PbI_3$ layers prepared by the laser evaporation technique to humid air at RH conditions of 40% and 75%. In our degradation experiments, no particular humidity control was made. For the degradation at 40% RH, the experiment was carried out in the winter season (January), whereas the degradation at 75% RH was implemented in the summer season (August). In our experiments, therefore, the RH was constant even during various measurements. For the degradation at 40% RH, however, the RH values fluctuate slightly by 5 % (i.e., 40 ± 5% RH).

During the degradation at 40% RH, various characterizations were made using real-time SE, XRD, scanning electron microscope (SEM), and atomic force microscopy (AFM). After the exposure of the sample to humid air, the sample was cut into three pieces to perform XRD, SEM, and AFM characterizations separately at similar timings. The SE measurement takes only 10 s and the SE spectra were obtained before the XRD measurements. After the 48-h air exposure (40% RH), the sample was further annealed at 100 °C for 30 min in vacuum to evaluate the thermal effect on the air-exposed layer. In particular, the $CH_3NH_3PbI_3 \cdot H_2O$ phase has been reported to be a metastable phase and shows a rapid structural change even at room temperature.[27] Thus, from the thermal stability of the air-exposed samples, the presence of the $CH_3NH_3PbI_3 \cdot H_2O$ phase can be discussed. In this study, the exposure of the $CH_3NH_3PbI_3$ samples to humid air was carried out under ordinary room illumination. Although the light illumination has been reported to accelerate $CH_3NH_3PbI_3$ degradation,[10,11,16,18,28] the precise control of light illumination was difficult during our measurement cycles.

For the characterization of $CH_3NH_3PbI_3$ samples without air exposure, we employed a $N_2$-filled glove box attached directly to the laser evaporation system to seal the $CH_3NH_3PbI_3$ samples inside a plastic bag. This plastic bag was opened inside a $N_2$-filled glove bag, in which the ellipsometry instrument was placed. This measurement was performed in a 0.7–5.2 eV energy range at an angle of incidence of 75° (J. A. Woollam M−2000XI). For ellipsometry characterization in air, we used another ellipsometer (J. A. Woollam, M−2000DI) that allows spectral measurements for up to 6.5 eV. The SE analysis for $CH_3NH_3PbI_3$ in a $N_2$ ambient has been reported elsewhere.[26]

The dielectric functions of the air-exposed/annealed layer on the ZnO/*c*−Si substrate were extracted directly using mathematical inversion[25] without incorporating a surface roughness contribution to simplify the SE analysis. In this SE analysis, we employed an optical model consisting of ambient/bulk layer (degraded layer)/interface layer (3 nm)/ZnO (50 nm)/$SiO_2$ (2 nm)/*c*−Si substrate. The interface layer corresponds to the surface roughness of the ZnO layer and the optical properties of this layer are calculated as a 50:50 vol.% mixture of the bulk layer and ZnO by applying the Bruggeman effective-medium approximation.[25,29] In the actual analysis, we first performed the SE fitting analysis in the transparent region by modeling the bulk-layer dielectric function using the Tauc-Lorentz model.[30] In the following analysis, the dielectric function of the bulk layer was extracted from the mathematical inversion using the effective bulk layer thickness estimated from the Tauc-Lorentz analysis.

On the other hand, the precise SE analysis was made for $PbI_2$ samples by incorporating a roughness contribution using a global error minimization scheme,[26] in which dielectric functions are determined accurately from multi-sample analysis.[31]

III. DFT CALCULATION

The DFT calculations were implemented using a plane-wave ultrasoft pseudopotential method (Advance/PHASE software). For the exchange-correlation functional, the generalized gradient approximation within the Perdew-Burke-Ernzerhof (PBE) scheme[32] has been applied. In this study, the simple PBE calculation is performed for $PbI_2$, $CH_3NH_3PbI_3 \cdot H_2O$ and $(CH_3NH_3)_4PbI_6 \cdot 2H_2O$ without incorporating the effect of spin-orbit coupling (SOC). In the DFT calculation of $CH_3NH_3PbI_3$, the experimental band gap ($E_g$) of 1.6 eV can be reproduced using PBE without considering the SOC interaction.[26,33-39] Nevertheless, it is now widely accepted that this result is caused by the cancellation of errors induced by two effects; i.e., (i) the underestimation of $E_g$ caused by the PBE functional and (ii) the overestimation of $E_g$ generated without considering SOC.[33-35] When the SOC interaction is incorporated into more sophisticated DFT calculations using the *GW* approximation[35,36] and hybrid functional,[37] the experimental $E_g$ of $CH_3NH_3PbI_3$ is reproduced. However, the band structures obtained from these advanced calculations are essentially similar to that deduced from the simple PBE calculation.[38] Thus, to simplify the DFT analysis, we performed the DFT calculations without incorporating SOC. The same approach is employed in recent DFT studies on $CH_3NH_3PbI_3$.[38,39]

The structural optimization of $PbI_2$ was performed using a 6 × 6 × 4 k mesh and a plane-wave cutoff energy of 950 eV until the atomic configuration converged to within 0.1 meV/Å. For the hydrate phases, the structural optimization was not made due to the large number of constituent atoms and the atomic configurations determined by XRD analyses were used for $CH_3NH_3PbI_3 \cdot H_2O$ (Ref. 27) and $(CH_3NH_3)_4PbI_6 \cdot 2H_2O$ (Ref. 40). The dielectric functions were calculated based on a method developed by Kageshima *et al*.[41]

## IV. RESULTS AND DISCUSSION

### A. Variation of structure during degradation

Figure 1 shows the SEM images of the laser evaporated layer obtained after (a) the 20-min air exposure and (b) the 48-h air exposure and the following thermal annealing. The RH condition for this experiment was 40%. The effect of the thermal annealing performed in Fig. 1(b) is quite weak and the layer structure changes only slightly after the annealing, as described below. In Fig. 1(a), the unintentional air exposure (20 min) was made for the preparation of the SEM measurement. As confirmed from Fig. 1(a), the $CH_3NH_3PbI_3$ layer with a thickness of 45 nm is quite smooth and the root-mean-square roughness of this layer, estimated by AFM, is only 4.6 nm.

It can be seen from Fig. 1(b) that the smooth $CH_3NH_3PbI_3$ structure changes completely after the prolonged air exposure and the air exposure of $CH_3NH_3PbI_3$ layers leads to the formation of a disk-like structure. Similar disk-like structures have been confirmed for conventional solution-processed $CH_3NH_3PbI_3$ after the humid air exposure.[8,13,18] Thus, the effect of humid air on $CH_3NH_3PbI_3$ is quite drastic. Closer inspection of the SEM image in Fig. 1(b) reveals that the platelet structure has a distorted hexagonal shape, which was attributed to the $PbI_2$ formation in earlier studies.[8,13] In fact, synthesized $PbI_2$ crystals show almost perfect hexagonal platelet structures,[42,43] which originate from the atomic arrangement of Pb and I.

Figure 2 shows the variation of the surface structure during the air exposure, obtained from SEM. In Fig. 2, the time of the air exposure is (a) 20 min, (b) 1 h, (c) 6 h, (d) 12 h, (e) 24 h, (f) 36h, (g) 48 h and (h) 48 h with thermal annealing for 40% RH and (i) 20 min for 75% RH. The SEM of Fig. 2(h) corresponds to the one shown in Fig. 1(b). The variation of the surface structure in Figs. 2(a)-(g) reveals (i) the formation of protrusion on the surface [Fig. 2(b)], (ii) the uniform coverage of the surface presumably by a new phase [Fig. 2(c)], (iii) the generation of pores on the surface [Fig. 2(d)], (iv) the enlargement of the pore structure [Fig. 2(e)], followed by (v) the gradual growth of the hexagonal platelet structure [Figs. 2(f) and (g)]. After the 36-h air exposure, the variation of the disk-like structure is rather negligible and the structure is maintained even after the 100-°C annealing, as confirmed from Fig. 2(h). The SEM characterization in Fig. 2 shows clearly that the change in the near-surface structure occurs rather rapidly with a time scale much faster than previously reported even at a low RH of 40%.

When the RH is higher (75% RH), on the other hand, the surface structure is quite different from that obtained at 40% RH. Specifically, the surface exposed to 75% RH is covered with a fiber-like structure with a short axis dimension of ~50 nm, as shown in Fig. 2(i). The formation of fiber-like structures at high RH conditions has also been confirmed previously[18,19] and thus the degradation of $CH_3NH_3PbI_3$ depends strongly on RH conditions.[8,18] It should be emphasized that, at 75% RH, the surface reaction is quite rapid and the mirror-like surface of the pristine $CH_3NH_3PbI_3$ layer becomes whitish within 20 s after the air exposure. Unfortunately, reliable optical characterization was quite difficult for this sample due to the extensive roughness and further analysis was not made for this sample. It should be noted that, when AFM measurements were performed using a conventional Si cantilever, we obtained quite smooth roughness even for the platelet structure shown in Fig. 2(g). In our case, therefore, the conventional AFM characterization cannot reproduce the SEM result and leads to the artificial flat surface due to the quite high aspect ratio in the platelet structure.

We determined the structure of the hexagonal platelet crystals formed at 40% RH using transmission electron microscopy (TEM). Figure 3 shows (a) the cross-sectional TEM image obtained after the air exposure at 40% RH and the thermal annealing and (b) the enlarged image of the same region. The TEM images were obtained from the same sample shown in Figs. 1(b) and 2(h). Quite surprisingly, we find that the platelet structure with a distorted hexagonal shape has a well-distinguished multilayer structure with a period of 6.5 ± 0.5 Å. The TEM image of Fig. 3(a) indicates that the multilayer stack is generated toward the short axis direction of the platelet structure and the direction of the multi-stack structure changes depending on the orientation of the platelet crystal. The multilayer structure was observed in all the magnified TEM images obtained from this sample, indicating the formation of the unique structure by the structural transformation in humid air.

To characterize the change in the crystal structure during the $CH_3NH_3PbI_3$ degradation, we performed the XRD analysis. Figure 4 summarizes the crystal structures of (a) $CH_3NH_3PbI_3$, (b) $PbI_2$, (c) $CH_3NH_3PbI_3 \cdot H_2O$ and (d) $(CH_3NH_3)_4PbI_6 \cdot 2H_2O$, which are assumed in the XRD analysis. $CH_3NH_3PbI_3$ (Ref. 26) and $PbI_2$ structures in Figs. 4(a) and (b) represent the DFT optimized structures, whereas the structures of $CH_3NH_3PbI_3 \cdot H_2O$ (Ref. 27) and $(CH_3NH_3)_4PbI_6 \cdot 2H_2O$ (Ref. 40), shown in Figs. 4(c) and (d), respectively, have been determined experimentally from the XRD analyses. In Fig. 4, the unit cell structures are denoted by black lines with the volume of each unit cell and the arrows represent the directions of the *a*, *b*, and *c* axes of the unit cell. For $CH_3NH_3PbI_3$, we assumed the pseudo-cubic structure shown in Fig. 4(a). The $CH_3NH_3PbI_3$ crystal has a unique three-dimensional structure, in which each $PbI_6$ octahedron is connected by corner I atoms. In the DFT result of Fig. 4(a), the C−N axis direction is almost parallel to the *a* axis, although other $CH_3NH_3^+$ orientations are also energetically possible.[33-35,37-39,44] The $PbI_2$ crystal has a two-dimensional structure, consisting of $PbI_6$ edge sharing octahedron in a hexagonal arrangement.[45-47] In particular, as shown in Fig. 4(b), the $PbI_2$ has the continuous stacking of a I–Pb–I unit and the separation of the Pb atom plane along the *c* axis is 7.0 Å, which correlates well with the period of 6.5 ± 0.5 Å observed in Fig. 3(b).

The hydrate phases shown in Figs. 4(c) and (d) are monoclinic crystals and have rather complicated structures. The $CH_3NH_3PbI_3 \cdot H_2O$ phase has a one-dimensional $PbI_6$ structure[27] and the cross-section of the one-dimensional crystal is consisting of two $PbI_6$ octahedron. In $(CH_3NH_3)_4PbI_6 \cdot 2H_2O$, on the other hand, the $PbI_6$ octahedron is completely separated each other and can be considered as a zero dimensional structure.[48-51] The configurations of $CH_3NH_3^+$ and $H_2O$ molecules in these hydrate phases can be interpreted by the strong interactions of $H_2O$ with (i) I atoms (O–H⋯I) and (ii) $CH_3NH_3^+$ (N–H⋯O).[7,48] Specifically, the polarity of $H_2O$ molecules generates the hydrogen bonding of $O(\delta^-)$–$H(\delta^+)$⋯$I(\delta^-)$ and $N(\delta^-)$–$H(\delta^+)$⋯$O(\delta^-)$, where $\delta^+$ and $\delta^-$ represent the positive and negative partial charges, respectively, which are determined essentially by the electronegativity. As a result, O−H bonds are oriented in the direction of the I atoms, whereas a N−H⋯O($H_2$)⋯H−N local structure is generated between $CH_3NH_3^+$ and $H_2O$.[48]

Figure 5 shows (a) the XRD spectra obtained experimentally during the degradation of the $CH_3NH_3PbI_3$ layer at 40% RH and (b) the XRD spectra calculated assuming the random orientation of the crystal structures shown in Fig. 4. The XRD pattern obtained from the CH$_3$NH$_3$PbI$_3$ with the minimum air exposure (20 min) is consistent with the XRD spectrum of the cubic phase,[52] although the XRD analysis of CH$_3$NH$_3$PbI$_3$ single crystals reveals that the tetragonal phase is the most stable phase at room temperature.[52,53] In this XRD spectrum, a PbI$_2$ peak at $2\theta = 12.7^\text{o}$ is quite weak, confirming that the generation of the PbI$_2$ secondary phase is negligible in the CH$_3$NH$_3$PbI$_3$ layer. A similar XRD spectrum has also been reported for a CH$_3$NH$_3$PbI$_3$ layer fabricated by a conventional evaporation process.[54]

During the degradation in humid air, the variation of the XRD pattern is quite small even after the 12-h air exposure, although the SEM measurements in Fig. 2 confirm the large structural change in the near-surface region. Thus, XRD is rather insensitive to the near-surface structural change, as the XRD intensities are average out by the whole layer. After the 24-h exposure, the intensities of PbI$_2$ diffraction peaks at 12.7$^\text{o}$, 38.7$^\text{o}$, and 52.4$^\text{o}$ increase notably, indicating the generation of the PbI$_2$ phase after the degradation. Quite interestingly, the 48-h exposure leads to the complete disappearance of the CH$_3$NH$_3$PbI$_3$ diffraction peaks at 14.1$^\text{o}$ and 43.2$^\text{o}$, respectively, even though the XRD peaks at 28.4$^\text{o}$ and 58.8$^\text{o}$ change only slightly. After the air exposure ($t \geq 24$ h), the XRD peaks at 28.4$^\text{o}$ and 58.8$^\text{o}$ shift slightly to 28.2$^\text{o}$ and 58.6$^\text{o}$, and the half width of the 28.2$^\text{o}$ diffraction peak increases by a factor of three, compared with the original peak at 28.4$^\text{o}$. However, the shifted XRD peaks at 28.2$^\text{o}$ and 58.6$^\text{o}$, observed after the 24-h exposure, are not reproduced well in the XRD spectra calculated assuming the (CH$_3$NH$_3$)$_4$PbI$_6$·2H$_2$O, CH$_3$NH$_3$PbI$_3$·H$_2$O and PbI$_2$ structures. As indicated in Fig. 5(b), the $(21\bar{2})$ diffraction of the CH$_3$NH$_3$PbI$_3$·H$_2$O crystal closely matches with the diffraction peak at 28.2$^\text{o}$. In Fig. 5(a), however, the 100-$^\text{o}$C thermal annealing does not change the overall XRD spectrum. Thus, the structure formed after the prolonged air exposure ($t \geq 48$ h) is thermally stable and is not likely to be thermally unstable CH$_3$NH$_3$PbI$_3$·H$_2$O (Ref. 27). On the other hand, it is obvious that the variation of the XRD spectrum in Fig. 5(a) cannot be explained by the simple phase transformation from CH$_3$NH$_3$PbI$_3$ to PbI$_2$. It should be noted that the diffraction at $2\theta = 28.2^\text{o}$ corresponds to a lattice plane separation of $d = 3.16$ Å, which is equivalent to a Pb–I bond length in CH$_3$NH$_3$PbI$_3$ shown in Fig. 4(a).

**B. Variation of optical properties during degradation**

To reveal the change of the optical properties during the phase transformation in humid air, we have applied real-time SE. Nevertheless, due to the quite complicated surface morphology formed during the phase transformation, the precise SE analysis has been quite difficult. Thus, we performed a rather simple SE analysis by assuming no roughness contribution, as described above. It is known quite well that the presence of surface roughness alters the amplitude of the dielectric function significantly particularly in a high energy region.[25,29] However, the energies of transition peaks are not influenced strongly by the roughness component[25,29] and the qualitative analysis can still be made. Moreover, the dimension of the platelet crystals is in a range of 20−100 nm, which is smaller than wavelengths of SE probe light. Accordingly, the average optical properties and the thickness of the bulk layer can be obtained to some extent from the above SE analysis.

Figure 6 shows (a) the variation of the $\varepsilon_2$ spectrum during the air exposure at 40% RH and (b) the $\varepsilon_2$ spectra of (a) in a selected energy region. In Fig. 6(a), each $\varepsilon_2$ spectrum is shifted by −1 for clarity. The $\varepsilon_2$ spectra of the pristine $CH_3NH_3PbI_3$ layer obtained in a $N_2$ ambient and the $PbI_2$ layer reported in Ref. 26 are also shown in Fig. 6(a). The transition energies denoted by $E_n$ ($n \leq 2$) in Fig. 6(a) represent the critical point energies of $CH_3NH_3PbI_3$. These optical transitions have been assigned to the direct semiconductor-type transitions at the R ($E_0 = 1.61$ eV), M ($E_1 = 2.53$ eV), and X ($E_2 = 3.24$ eV) points in the pseudo-cubic Brillouin zone.[26] In the spectrum obtained after the 20-s air exposure, we observed an additional $\varepsilon_2$ peak at $E_3 = 5.65$ eV.

The $\varepsilon_2$ spectrum basically represents light absorption properties and the $E_0$ transition of $CH_3NH_3PbI_3$ shows the interband transition at the $E_g$. As confirmed from Fig. 6, the amplitude of the $E_0$ transition reduces gradually with the exposure time and the $E_0$ transition disappears almost completely at $t \geq 24$ h. During the degradation, the broad $E_1$ transition of $CH_3NH_3PbI_3$ changes gradually into a sharp transition, which is consistent with the phase transformation from $CH_3NH_3PbI_3$ to $PbI_2$. The sharp transition peak at 2.51 eV in $PbI_2$ has been assigned to the excitonic transition[46,47] and the appearance of a $\varepsilon_2$ peak at 2.97 eV at $t \geq 16$ h also supports the presence of the $PbI_2$ phase. In fact, the spectral feature of $E < 3$ eV in the degraded layer can be represented almost perfectly by that of $PbI_2$. After the degradation, therefore, the absorption onset shifts from 1.6 eV in $CH_3NH_3PbI_3$ to 2.4 eV in $PbI_2$ and the prepared layer becomes transparent due to the strong absorption loss in the visible region. Similar results have already been obtained in earlier studies, leading to the conclusion that $CH_3NH_3PbI_3$ is degraded into the $PbI_2$ phase upon air exposure.[8,10,11,16,17]

Nevertheless, the spectral feature in the UV region ($E > 3$ eV) obtained after the degradation differs significantly from that of $PbI_2$. In Fig. 6, the $PbI_2$ transition peaks at 2.51, 2.97, 3.26, 3.90, and 4.33 eV are indicated by arrows and all of these peaks can be confirmed in the air-exposed layer ($t \geq 16$ h ). However, the degraded phase shows a strong transition peak at 3.41 eV, which cannot be explained by the $PbI_2$ formation. Since the visible light absorption disappears in the degraded layer, this peak cannot be attributed to the interband transition in the $CH_3NH_3PbI_3$ phase. Thus, the presence of the strong 3.41-eV absorption strongly supports the formation of a new phase, other than $CH_3NH_3PbI_3$ and $PbI_2$. So far, $CH_3NH_3PbI_3 \cdot H_2O$ (Ref. 19) and $(CH_3NH_3)_4PbI_6 \cdot 2H_2O$ (Refs. 50, 51) crystals are reported to show intense transition peaks at 3.09 eV and 3.33 eV, respectively. Thus, the intense transition peak at 3.41 eV suggests the generation of a $CH_3NH_3PbI_3$ hydrate crystal, as discussed later. On the other hand, the gradual reduction of the $E_3$ peak intensity, observed during the degradation, shows that this peak originates from the interband transition in $CH_3NH_3PbI_3$. The result of Fig. 6 also indicates that the thermal annealing at 100 $^oC$ does not modify the spectrum, confirming that the degraded phase is thermally stable.

To determine the optical properties of the hydrated crystal phases, we implemented the DFT calculations assuming the crystal structures shown in Fig. 4. Figure 7 shows the calculated band structures and partial density of states (pDOS) of (a) $PbI_2$, (b) $CH_3NH_3PbI_3 \cdot H_2O$ and (c) $(CH_3NH_3)_4PbI_6 \cdot 2H_2O$, together with (d) the Brillouin zones of the $PbI_2$ crystal (hexagonal) and the hydrate crystals (monoclinic). $PbI_2$ and $CH_3NH_3PbI_3 \cdot H_2O$ are direct transition semiconductors and the direct transitions occur at the A point $(PbI_2)$[45,46] and the Z point $(CH_3NH_3PbI_3 \cdot H_2O)$. In the case of $(CH_3NH_3)_4PbI_6 \cdot 2H_2O$, the position of the valence band maximum (VBM) is different from that of the conduction band minimum (CBM) and the transition is indirect. However, the band dispersion is quite flat and the direct transition energy at the A point is almost equivalent to the transition energy of the indirect gap. Thus, $(CH_3NH_3)_4PbI_6 \cdot 2H_2O$ is a pseudo-direct semiconductor.

The pDOS contributions in $PbI_2$ and the hydrate phases are essentially similar to those of $CH_3NH_3PbI_3$. Specifically, the CBM of these crystals consists of Pb $6p$ state, whereas the VBM is dominated by I $5p$. In $PbI_2$ and $CH_3NH_3PbI_3 \cdot H_2O$, there are small contributions of Pb $6s$ near the VBM, whereas the distribution of Pb $6s$ at the VBM increases in $(CH_3NH_3)_4PbI_6 \cdot 2H_2O$ having an isolated structure of the $PbI_6$ octahedron. The $CH_3NH_3^+$ and $H_2O$ in the hydrate phases exhibit weak interactions with the VBM and CBM, generating localized states in the deeper region from the VBM, as confirmed also for $CH_3NH_3PbI_3$.[38,55] These results indicate that the optical transitions in the hydrates are determined predominantly by the $PbI_6$ component. However, the $E_g$ of these crystals varies strongly with the arrangement of the $PbI_6$ octahedron and the calculated $E_g$ increases from 2.18 eV in $PbI_2$ to 2.52 eV in $CH_3NH_3PbI_3 \cdot H_2O$ and 3.53 eV in $(CH_3NH_3)_4PbI_6 \cdot 2H_2O$.

Figure 8 shows the calculated dielectric functions of (a) $PbI_2$, (b) $CH_3NH_3PbI_3 \cdot H_2O$, (c) $(CH_3NH_3)_4PbI_6 \cdot 2H_2O$, together with (d) isotropic dielectric functions [$(\varepsilon_x + \varepsilon_y + \varepsilon_z)/3$] of $PbI_2$, $CH_3NH_3PbI_3 \cdot H_2O$ and $(CH_3NH_3)_4PbI_6 \cdot 2H_2O$. In Fig. 8, the dielectric functions denoted as $\varepsilon_x$, $\varepsilon_y$ and $\varepsilon_z$ represent those when the polarization state is parallel to orthogonal $x$, $y$ and $z$ axes and, in the case of a cubic structure, $x$, $y$, and $z$ axes correspond to the $a$, $b$, and $c$ axes of the unit cell, respectively. The experimental dielectric functions of $PbI_2$ (Fig. 6) and $CH_3NH_3PbI_3 \cdot H_2O$ (Ref. 19) are also shown in Fig. 8. For these materials, the DFT results show rather poor agreement with the experimental results. In particular, for $CH_3NH_3PbI_3 \cdot H_2O$, the DFT calculation does not reproduce a strong $\varepsilon_2$ peak observed experimentally at 3.09 eV. The earlier study showed that this transition is excitonic[19] and this optical transition cannot be expressed in our simple DFT calculations performed without considering excitonic effects. In Fig. 8(c), the spectra of absorption coefficient ($\alpha$) for $(CH_3NH_3)_4PbI_6 \cdot 2H_2O$, measured at room temperature (RT)[50,51] and 2 K (Ref. 49), are also indicated. It can be seen that the $\varepsilon_y$ contribution shows a similar trend with the $\alpha$ spectrum at 2 K. The $\alpha$ spectrum at 2 K also shows quite good correlation with the pDOS distribution near the VBM [Fig. 7(c)]. Accordingly, our simple DFT calculation using PBE appears to reproduce the experimental result of $(CH_3NH_3)_4PbI_6 \cdot 2H_2O$ well, although the calculated transition energies are slightly different from those observed experimentally. At RT, the experimental $\alpha$ spectrum exhibits a transition peak at 3.33 eV, which can be interpreted as the direct interband transition in Fig. 7(c).

Since the hydrate crystals formed by the degradation are expected to have random polycrystalline structures, we estimated the isotropic optical properties from $(\varepsilon_x + \varepsilon_y + \varepsilon_z)/3$ and the calculation results for $PbI_2$ and the hydrate crystals are compared in Fig. 8(d). The experimental dielectric function obtained after the 24-h air exposure is also shown in this figure. It can be seen that the absorption edge shifts toward higher energy as the two-dimensional octahedron structure of $PbI_2$ changes into more isolated $PbI_6$ structures having lower dimensions. In other word, the interaction between the neighboring $PbI_6$ octahedrons contributes to reduce $E_g$.

Based on the above results, we propose that the strong optical transition observed at 3.41 eV in Fig. 6 originates from the isolated (zero-dimensional) $PbI_6$ structure present in a $CH_3NH_3PbI_3$ hydrate phase. The peak position of 3.41 eV observed in the air-exposed sample agrees quite well with the reported peak position of 3.33 eV in $(CH_3NH_3)_4PbI_6 \cdot 2H_2O$.[50,51] The XRD peaks at $2\theta = 28.2^o$ and $58.6^o$ observed after the complete phase transformation indicate the presence of a different phase other than $PbI_2$ and are consistent with the generation of the hydrate phase. However, the XRD spectrum obtained from the degraded layer does not match with that of $(CH_3NH_3)_4PbI_6 \cdot 2H_2O$ and the local arrangement is considered to be different. Elucidation of the exact crystal structure in the hydrate phase requires further study.

So far, many SE studies have been reported for $CH_3NH_3PbI_3$.[19,56-58] In these studies, however, the particular control of the measurement environment was not made and a broad $\varepsilon_2$ peak at 3.4 eV, which is similar to that observed in the degraded layer, has been reported. Our result in Fig. 6 indicates that the broad $\varepsilon_2$ transition at 3.4 eV consists of two interband transitions originating from the $PbI_2$ peak at 3.26 eV and the hydrate peak at 3.41 eV.

## C. Degradation mechanism of $CH_3NH_3PbI_3$

Figure 9 summarizes the real-time SE results obtained for the $CH_3NH_3PbI_3$ degradation at 40% RH. In this figure, (a) effective thickness of the degraded layer, (b) $E_2$ peak position and (c) $\varepsilon_2$ value at 2.0 eV are plotted as a function of the exposure time to humid air. The effective thickness represents the thickness of the single bulk layer assumed in the SE analysis. The time evolution of the effective thickness indicates the rapid thickness reduction by 6 nm in the initial stage ($t \leq 3$ h), followed by the gradual decrease that continues up to 28 h. As mentioned earlier, in the SE analysis, the effective thickness was estimated by assuming a flat layer for the air-exposed layer. As confirmed from Fig. 2, the layer structure is rather flat at $t \leq 24$ h and the effective thickness can be determined without significant errors. After $t > 24$ h, the SEM measurement shows the formation of the platelet structure, but the effective thickness changes only slightly. We note that the SE result at $t > 24$ h may include relatively large errors due to the large structural non-uniformity.

For the degradation of $CH_3NH_3PbI_3$ in humid air, the formation processes of $PbI_2$ (Refs. 9-14), $CH_3NH_3PbI_3 \cdot H_2O$ (Ref. 19) and $(CH_3NH_3)_4PbI_6 \cdot 2H_2O$ (Ref. 8) have been proposed to occur by

$$CH_3NH_3PbI_3 \rightarrow PbI_2 + CH_3NH_3I \text{ (or } CH_3NH_2 + HI) \qquad (1)$$
$$CH_3NH_3PbI_3 + H_2O \rightarrow CH_3NH_3PbI_3 \cdot H_2O \qquad (2)$$
$$4(CH_3NH_3PbI_3) + 2H_2O \rightarrow (CH_3NH_3)_4PbI_6 \cdot 2H_2O + 3PbI_2 \qquad (3)$$

It has been reported that the reaction (2) is a reversible reaction and, under vacuum, $CH_3NH_3PbI_3$ is formed from $CH_3NH_3PbI_3 \cdot H_2O$ by $H_2O$ desorption.[19] From the volumes of the unit cells shown in Fig. 4, it can be understood that the reduction of the layer volume (or the layer thickness) occurs only by the reaction (1) and, in the cases of the reactions (2) and (3), the volumes expand by the incorporation of $H_2O$ molecules. Thus, the SE result in Fig. 9(a) indicates that the effective thickness reduces rapidly by the PbI$_2$ formation and the following desorption of CH$_3$NH$_3$I-related species. This is consistent with the appearance of the sharp absorption feature of PbI$_2$ at 2.51 eV in Fig. 6.

In Fig. 9(b), the $E_2$ peak position extracted from the $\varepsilon_2$ spectra in Fig. 6 is shown. Quite interestingly, the $E_2$ peak shift correlates almost perfectly with the thickness reduction in Fig. 9(a). As discussed above, we attributed the $E_2$ peak at 3.41 eV to the formation of a CH$_3$NH$_3$PbI$_3$ hydrate crystal. Thus, the observed shift of the $E_2$ peak indicates the gradual phase transformation from CH$_3$NH$_3$PbI$_3$ to the hydrate crystal, while the rapid reduction in the layer thickness in Fig. 9(a) indicates the PbI$_2$ formation. These results present the important evidence that the CH$_3$NH$_3$PbI$_3$ degradation at 40% RH proceeds by two competing formation reactions of the PbI$_2$ and the hydrate phases. In the initial degradation process ($t \leq 3$ h), on the other hand, the change of the $\varepsilon_2$ spectrum in the visible region is quite small, whereas the $E_2$ peak shows a distinct peak shift. This can be interpreted by the formation of the hydrate phase in the surface region. In particular, the light absorption in the UV region is more sensitive to the near-surface structure as the penetration depth ($d_p$) of light is quite small due to high $\alpha$, and we obtain $d_p = 19$ nm at 3.4 eV.[26]

In the hydrate formation by the reaction (3), the total crystal volume expands by 1%, while the volume reduction in the reaction (1) is 52%. The result of Fig. 9(a) shows that the final thickness (28 nm) is 62% of the initial thickness (45 nm). By considering the unit-cell volume change and the thickness reduction after the degradation, we roughly estimated the hydrate volume fraction to be ~15%, assuming that the hydrate structure is (CH$_3$NH$_3$)$_4$PbI$_6$·2H$_2$O. The observed intense XRD peak at 28.2° and strong transition peak at 3.41 eV also suggest a relatively high volume fraction of the hydrate phase in the degraded layer. At a high RH condition of 80%, on the other hand, the degradation is reported to occur through the reaction (2) and the layer thickness increases slightly during the phase change.[19]

In Fig. 9(c), the $\varepsilon_2$ values at 2.0 eV in Fig. 6(b) are plotted and the light absorption at 2.0 eV decreases gradually during the phase formation. However, the variation in the visible light absorption is quite insensitive to the rapid thickness reduction and the hydrate formation in the surface region, confirmed in Figs. 9(a) and (b). Accordingly, when conventional optical spectroscopy in the visible region and XRD are employed, only the bulk component of samples is characterized and, in this case, the nominal degradation takes place very slowly particularly when thicker CH$_3$NH$_3$PbI$_3$ layers are evaluated.

We have further investigated the atomic configuration of the platelet structure in Fig.

3(b). Figure 10 shows the layer structure of $PbI_2$. As mentioned earlier, $PbI_2$ has a stacked-layer structure consisting of a I–Pb–I unit and the separation of the Pb atomic plane (7.0 Å) agrees with the observed interlayer distance of 6.5 ± 0.5 Å. In $PbI_2$, each I–Pb–I unit along the $c$ axis direction is combined by weak van der Waals interaction. Thus, compared with the $c$ axis, the growth rate of the $PbI_2$ crystal is much faster in the directions parallel to the $a$ and $b$ axes, which in turn leads to the formation of the platelet structure with a hexagonal shape. All the results obtained from SEM, XRD and SE undoubtfully indicate the presence of the $PbI_2$ phase.

It is obvious from Figs. 4 and 10 that the distinct multi-stack structure observed experimentally by TEM can best be explained by the $PbI_2$ formation. However, the XRD and SE results strongly indicate the presence of the hydrate phase. From the above results, we suggest that the platelet crystal formed after the $CH_3NH_3PbI_3$ degradation has non-negligible $(CH_3NH_3)_4PbI_6·2H_2O$-like local arrangements dispersed within the two-dimensional $PbI_2$ crystal. The distortion of the hexagonal structure, as observed in Fig. 1(b), can be explained by such a microscopically non-uniform structure. As a result, the $CH_3NH_3PbI_3$ degradation in humid air can be viewed as the transformation of the $PbI_6$-based crystal to a low dimensionality and the $PbI_6$ octahedron network changes from three dimensions ($CH_3NH_3PbI_3$) to two dimensions ($PbI_2$) or zero dimension [$(CH_3NH_3)_4PbI_6·2H_2O$], which alters the optical properties and shifts the absorption onset toward higher energies.

Figure 11 shows the proposed degradation mechanism of $CH_3NH_3PbI_3$ at 40% RH. The initial structure before the air exposure is shown in Fig. 11(a). After the air exposure, the desorption of $CH_3NH_3I$ species occurs by the reaction (1), which reduces the layer thickness by 4 nm within 1 h and generates the $PbI_2$ phase near the surface region. This thickness reduction corresponds to the phase transformation of a 8-nm-thick $CH_3NH_3PbI_3$ layer into a 4-nm-thick $PbI_2$ layer, if the volume change in Fig. 4 is taken into account. During the $CH_3NH_3I$ desorption, the hydrate formation also occurs by a reaction that is expected to be similar to the reaction (3). The $PbI_2$ and hydrate formation reactions are competing processes and occur simultaneously [Fig. 11(b)]. As confirmed from Figs. 9(a) and (b), the rates of the thickness reduction and the $E_2$ peak shift become slower at $t ≥ 3$ h. The SEM image obtained after the 1-h exposure [Fig. 2(b)] shows the formation of the protrusion or the nucleation of the $PbI_2$/hydrate phase, followed by the uniform coverage of the surface by this phase after the 6-h exposure [Fig. 2(c)]. These results suggest that the phase transformation near the surface is suppressed after 3 h due to the uniform coverage of the surface by the $PbI_2$/hydrate phase.

After the 3-h exposure, the formation of the hexagonal-shaped $PbI_2$/hydrate crystals occurs gradually. It should be emphasized that the diameter of the platelet crystals is ~100 nm [Figs. 1(b) and Figs. 2(f)-(h)] and the geometrical dimension is larger than the original thickness of the $CH_3NH_3PbI_3$ layer (45 nm). This indicates clearly that the large distorted hexagonal crystals are formed by the extensive diffusion process inside the bulk component. We speculate that the $PbI_2$ crystal grows toward the *a* and *b* axis direction by the diffusion of free Pb and I atoms, as illustrated in Fig. 10, and the hydrate phase is also formed within the $PbI_2$ crystal [Fig. 11(c)]. During this process, the pore structure, confirmed in the SEM images in Figs. 2(c)–(e), is generated on the surface as a result of the extensive diffusion. Unfortunately, the internal film structure during this process remains ambiguous, as the TEM observation is difficult for this intermediate structure. As shown in Fig. 11(d), the prolonged air exposure completely removes the $CH_3NH_3PbI_3$ phase, eliminating the light absorption in the visible region, and all the $CH_3NH_3PbI_3$ crystals are converted into the $PbI_2$/hydrate crystals.

It should be emphasized that the time scale of the $CH_3NH_3PbI_3$ degradation observed in this study at 40% RH (1 day) is much faster than those confirmed earlier using XRD and optical spectroscopy at a similar RH (1 month) [12,15,18] and correlates quite well with those observed in actual $CH_3NH_3PbI_3$ solar cells (1–2 days at 55% RH).[3] Accordingly, the degradation of hybrid perovskite solar cells in humid air is most likely caused by the rapid phase change of $CH_3NH_3PbI_3$ near the device interface region.

Finally, we note that the degradation behavior of $CH_3NH_3PbI_3$ may show slight changes depending on a sample preparation method and the resulting structure. In particular, solution-processed $CH_3NH_3PbI_3$ layers generally exhibit distinct grain boundary structures due to the formation of large polycrystalline grains with sizes of 300–500 nm.[1,57,59] It has been reported that, in the grain boundary region, the $PbI_2$-dominant structure is formed during thermal annealing,[59] which is adopted commonly as a post treatment process of various $CH_3NH_3PbI_3$ layers. In solution-processed $CH_3NH_3PbI_3$ layers having large grains, therefore, the degradation could also proceed by the reaction/diffusion at the grain boundaries. In fact, rapid degradation in the $CH_3NH_3PbI_3$ grain boundary region due to faster transport of $H_2O$ molecules along grain boundaries has been proposed.[19] Nevertheless, the formation of the hexagonal platelet crystals by the degradation is generally observed in solution-processed $CH_3NH_3PbI_3$ layers.[8,13,18] Thus, the final product formed after the humid air exposure is unaffected by the detailed structures of $CH_3NH_3PbI_3$.

## V. Summary

We have determined the degradation process of $CH_3NH_3PbI_3$ in humid air (40% RH) by real-time SE. The ultra-smooth surface of $CH_3NH_3PbI_3$ layers, prepared by a laser evaporation technique, allows us to identify the phase transformation that occurs near the surface region. The XRD characterization reveals that $CH_3NH_3PbI_3$ crystals change into $PbI_2$ crystals after the air exposure. However, a XRD peak at ~28.4$^o$, observed originally in the $CH_3NH_3PbI_3$ phase, shows minor variation upon air exposure, and we attributed this XRD peak to the formation of a $CH_3NH_3PbI_3$ hydrate phase. The formation of the hydrate phase in humid air is supported strongly by the presence of a distinct $\varepsilon_2$ peak observed at 3.41 eV. The DFT analysis confirms that an isolated $PbI_6$ octahedron in the hydrate phase induces a sharp optical transition in a similar energy region. From real-time SE, we find that the $CH_3NH_3PbI_3$ degradation at 40% RH can be explained by two parallel reactions of (i) $CH_3NH_3I$ desorption that leads to the $PbI_2$ formation ($CH_3NH_3PbI_3 \rightarrow PbI_2 + CH_3NH_3I$) and (ii) the formation of a $CH_3NH_3PbI_3$ hydrate crystal by $H_2O$ incorporation. After the long air exposure, the $CH_3NH_3PbI_3$ layer is transformed completely into distorted hexagonal platelet crystals, which show a well-defined multilayer structure with an interspacing of 6.5 ± 0.5 Å. The formation of the platelet structure is explained by the strong internal diffusion of Pb and I atoms. The XRD and optical characterization in the visible region are quite insensitive to the near-surface structural change and the phase transformation in the surface region occurs quite rapidly with a time scale much faster than those reported in earlier studies.

**Figure captions**

FIG. 1. SEM images of the laser evaporated layer obtained after (a) the 20-min air exposure and (b) the 48-h air exposure and the following thermal annealing at 100 $^{o}$C (30 min). The RH condition of this experiment was 40%. The SEM image of (b) was measured at a more inclined angle.

FIG. 2. Variation of the surface structure obtained from SEM with the air-exposure time of (a) 20 min, (b) 1 h, (c) 6 h, (d) 12 h, (e) 24 h, (f) 36h, and (g) 48 h for the degradation at 40% RH and (i) 20 min for the degradation at 75% RH. The SEM image of (h) was obtained after the 48-h air exposure, followed by 100 $^{o}$C annealing, and this SEM image was obtained from the identical sample shown in Fig. 1(b).

FIG. 3. (a) Cross-sectional TEM image obtained after the air-exposure at 40% RH and thermal annealing and (b) the enlarged image of the same region. The TEM images were obtained from the sample shown in Figs. 1(b) and 2(h).

FIG. 4. Crystal structures of (a) $CH_3NH_3PbI_3$, (b) $PbI_2$, (c) $CH_3NH_3PbI_3 \cdot H_2O$ and (d) $(CH_3NH_3)_4PbI_6 \cdot 2H_2O$. The crystal structures in (c) and (d) are those reported in Ref. 27 and Ref. 40, respectively. The unit cell structures are denoted by black lines and the volume ($V$) of the unit cell for each structure is also indicated. The arrows represent the directions of the $a$, $b$, and $c$ axes of the unit cell. The actual compositions of the unit cells in (c) and (d) are $2(CH_3NH_3PbI_3 \cdot H_2O)$ and $2[(CH_3NH_3)_4PbI_6 \cdot 2H_2O]$, respectively.

FIG. 5. (a) XRD spectra obtained experimentally during the degradation of the $CH_3NH_3PbI_3$ layer at 40% RH and (b) the XRD spectra calculated from the crystal structures of $CH_3NH_3PbI_3$, $(CH_3NH_3)_4PbI_6 \cdot 2H_2O$, $CH_3NH_3PbI_3 \cdot H_2O$ and $PbI_2$ shown in Fig. 4. The diffraction peaks originating from the underlying ZnO layer and the $c$−Si substrate are also indicated in (a).

FIG. 6. (a) Variation of the $\varepsilon_2$ spectrum during the air exposure at 40% RH and the following thermal annealing at 100 $^{o}$C and (b) the $\varepsilon_2$ spectra of (a) in a selected energy region. The dielectric functions of $CH_3NH_3PbI_3$ in $N_2$ and $PbI_2$ correspond to those reported in Ref. 26, but the amplitude of the $\varepsilon_2$ spectrum is reduced to half for $PbI_2$. In (a), each $\varepsilon_2$ spectrum is shifted by −1 for clarity. The $E_n$ ($n \leq 3$) indicates the transition energy in $CH_3NH_3PbI_3$.

FIG. 7. Band structures and pDOS of (a) PbI$_2$, (b) CH$_3$NH$_3$PbI$_3$·H$_2$O and (c) (CH$_3$NH$_3$)$_4$PbI$_6$·2H$_2$O, together with (d) the Brillouin zones of the PbI$_2$ (hexagonal) and hydrate (monoclinic) phases. For pDOS, only the contributions of Pb 6$s$, Pb 6$p$ and I 5$p$ are shown, as the pDOS values for C, N, O, and H atoms are quite small in the energy region.

FIG. 8. Calculated dielectric functions of (a) PbI$_2$, (b) CH$_3$NH$_3$PbI$_3$·H$_2$O and (c) (CH$_3$NH$_3$)$_4$PbI$_6$·2H$_2$O, together with (d) the isotropic dielectric functions [($\varepsilon_x + \varepsilon_y + \varepsilon_z$)/3] of PbI$_2$, CH$_3$NH$_3$PbI$_3$·H$_2$O and (CH$_3$NH$_3$)$_4$PbI$_6$·2H$_2$O. The dielectric functions denoted as $\varepsilon_x$, $\varepsilon_y$ and $\varepsilon_z$ represent those when the polarization state is parallel to orthogonal $x$, $y$ and $z$ axes and, in the case of a cubic structure, $x$, $y$, and $z$ axes correspond to the $a$, $b$, and $c$ axes of the unit cell, respectively. In (a) and (b), the experimental dielectric functions of PbI$_2$ (Ref. 26) and CH$_3$NH$_3$PbI$_3$·H$_2$O (Ref. 19) are shown. In (c), the experimental $\alpha$ spectra of the (CH$_3$NH$_3$)$_4$PbI$_6$·2H$_2$O phase at RT (Ref. 50) and 2 K (Ref. 49) are also shown.

FIG. 9. Real-time SE results obtained for the CH$_3$NH$_3$PbI$_3$ degradation at 40% RH: (a) effective thickness of the degraded layer, (b) $E_2$ peak position and (c) $\varepsilon_2$ value at 2.0 eV, plotted as a function of the exposure time to humid air. The effective thickness represents the thickness of the single bulk layer assumed in the SE analysis.

FIG. 10. Layer structure of PbI$_2$. The arrows represent the directions of the $a$, $b$, and $c$ axes of the unit cell.

FIG. 11. Proposed degradation mechanism of CH$_3$NH$_3$PbI$_3$ at 40% RH: (a) initial CH$_3$NH$_3$PbI$_3$ structure, (b) the formation of the PbI$_2$/hydrate phase near the surface, (c) the growth of the PbI$_2$/hydrate crystal by internal diffusion and (d) the formation of the platelet crystal after the prolonged air exposure. The dotted line indicates the position of CH$_3$NH$_3$PbI$_3$ surface in (a).

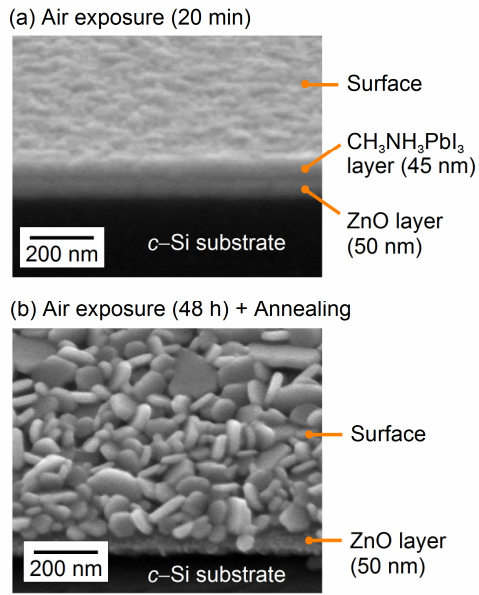

**Fig. 1.**

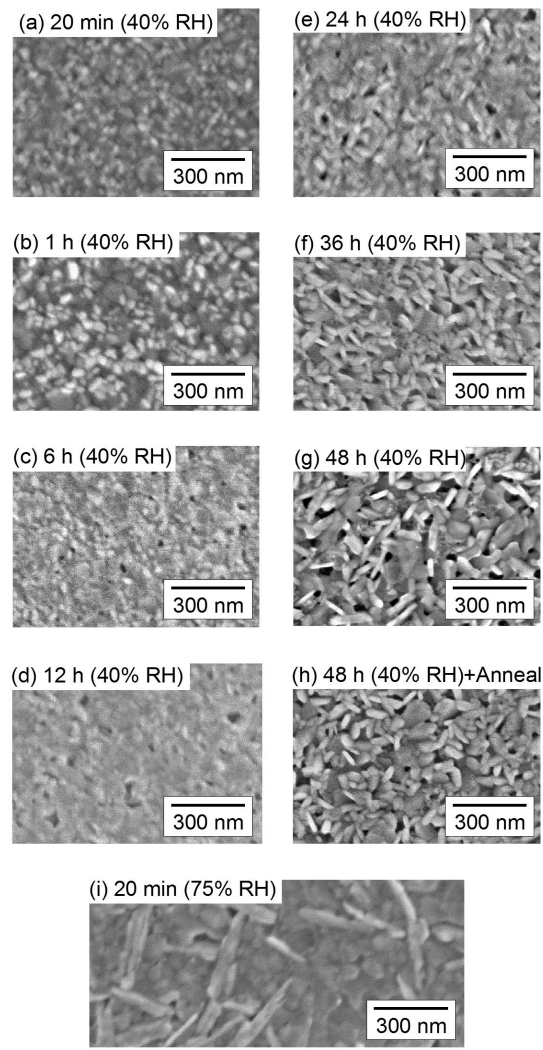

**Fig. 2.**

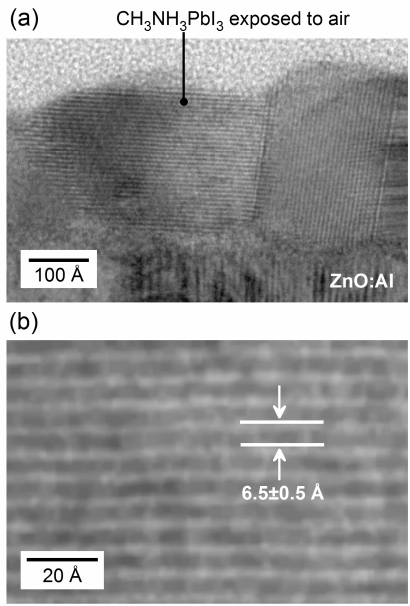

(a) CH$_3$NH$_3$PbI$_3$ exposed to air

100 Å

ZnO:Al

(b)

6.5±0.5 Å

20 Å

**Fig. 3.**

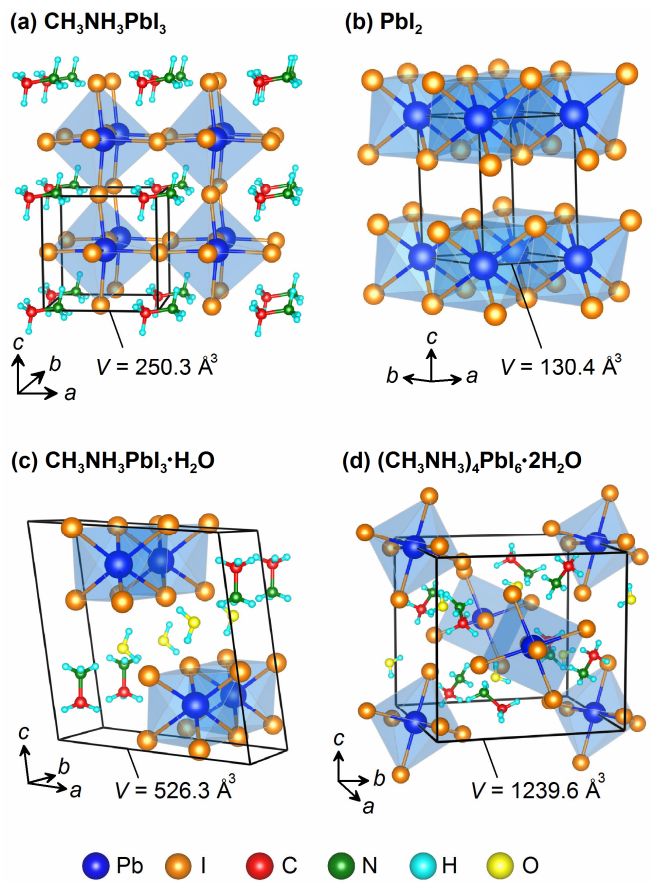

(a) CH$_3$NH$_3$PbI$_3$ — $V$ = 250.3 Å$^3$
(b) PbI$_2$ — $V$ = 130.4 Å$^3$
(c) CH$_3$NH$_3$PbI$_3$·H$_2$O — $V$ = 526.3 Å$^3$
(d) (CH$_3$NH$_3$)$_4$PbI$_6$·2H$_2$O — $V$ = 1239.6 Å$^3$

Pb   I   C   N   H   O

**Fig. 4.**

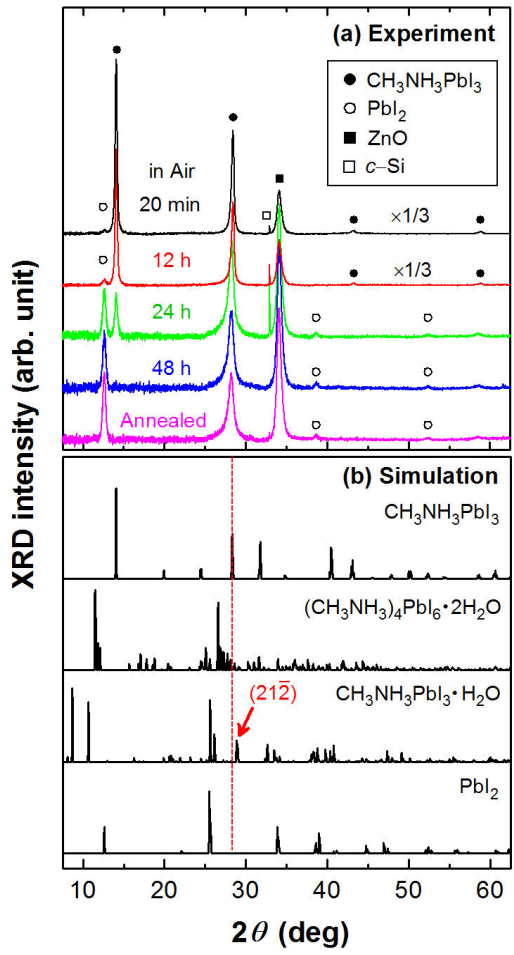

Fig. 5.

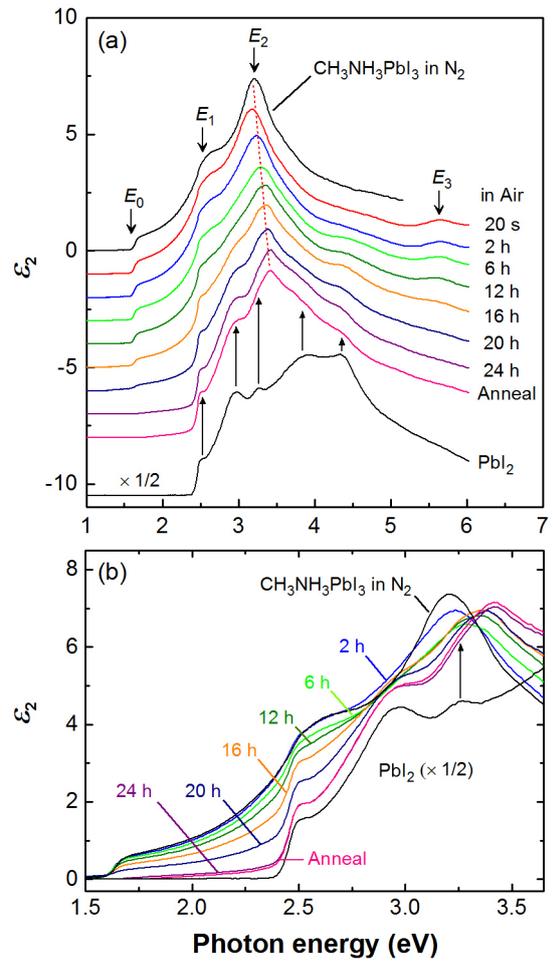

Fig. 6.

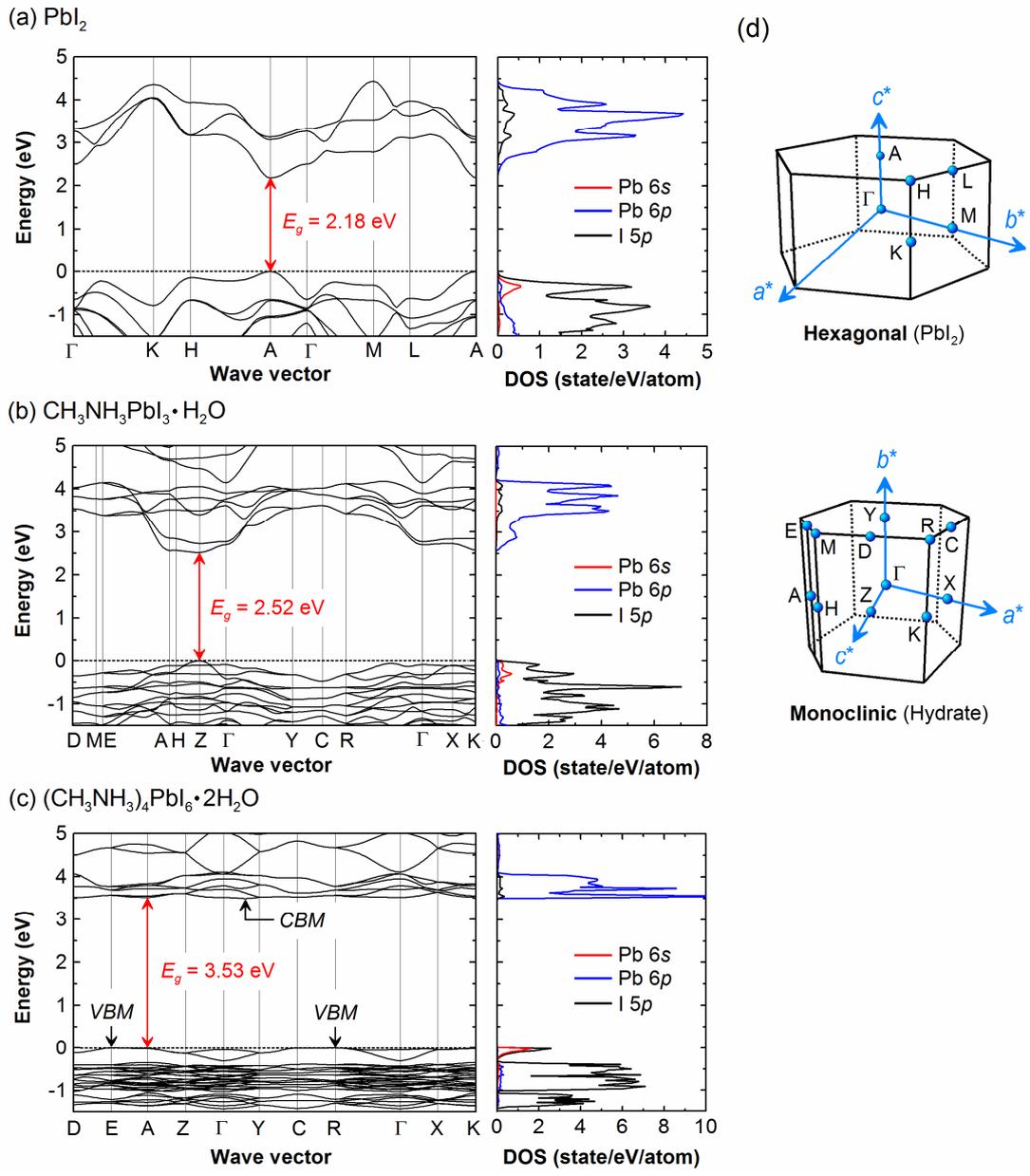

**Fig. 7.**

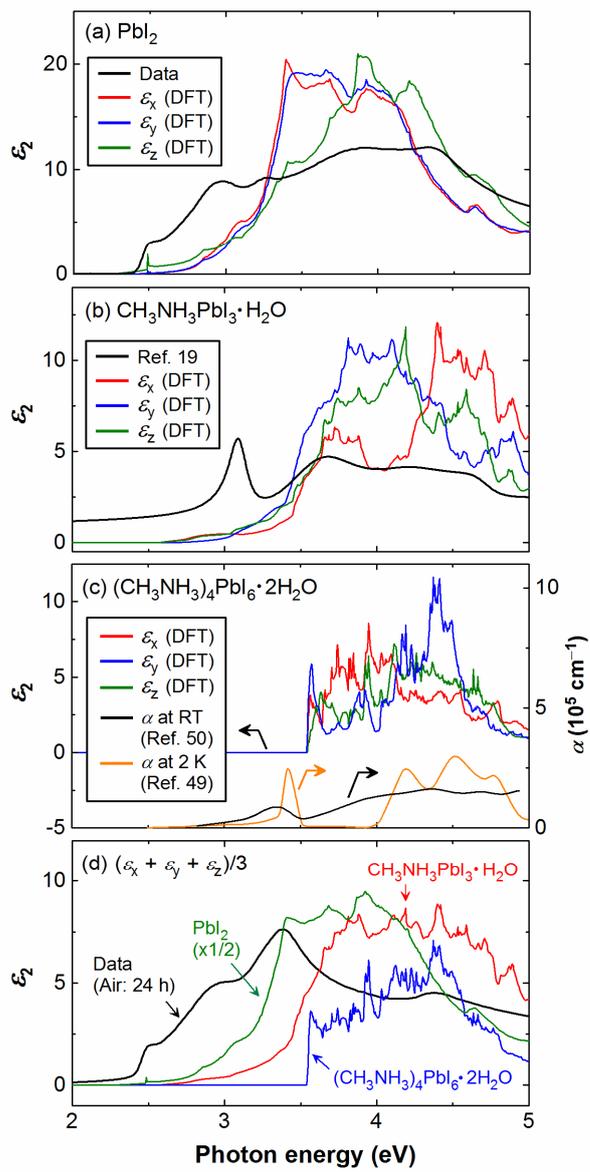

**Fig. 8.**

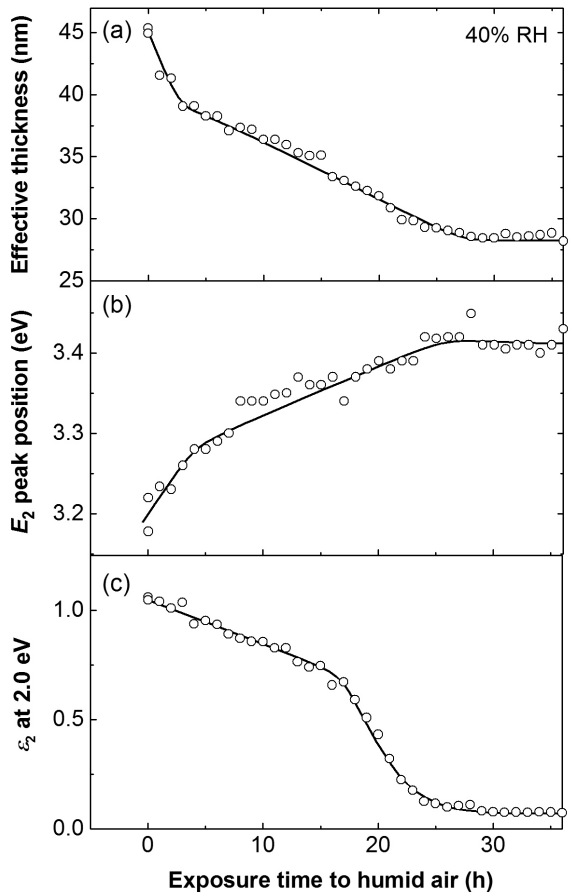

**Fig. 9.**

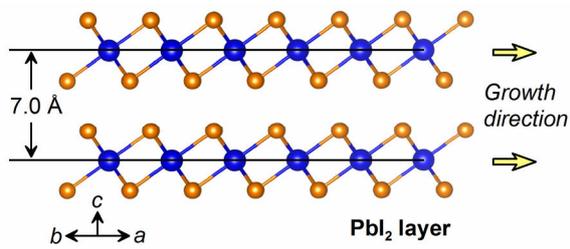

**Fig. 10.**

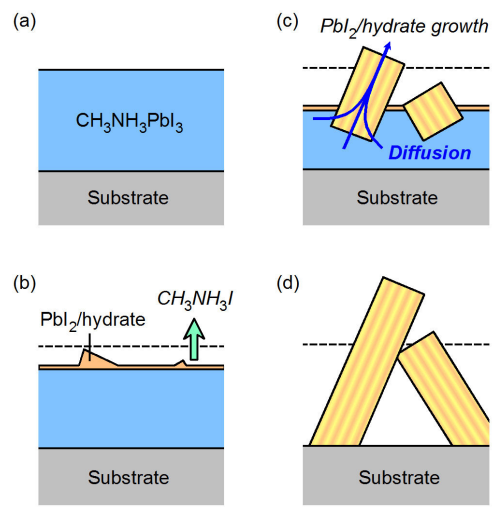

**Fig. 11.**